\documentclass[structabstract]{aa}

\usepackage{graphicx}
\usepackage{txfonts}
\usepackage{natbib}

\begin{document}


\title {Perspectives of current-layer diagnostics \\ in solar flares}
\author{A.V.~Oreshina \and I.V.~Oreshina}
\institute{P.K.~Sternberg Astronomical Institute,
M.V.~Lomonosov Moscow State University,
Universitetskii prospekt, 13, Moscow, 119991, Russia\\
\email{avo@sai.msu.ru}}

\date{Received 7 September 2012; accepted 21 July 2013}

\abstract
{A reconnecting current layer is a `heart' of a solar
flare, because it is a place of magnetic-field energy release.
However there are no direct observations of these layers.}
{The aim of our work is to understand why we actually
do not directly observe current layers
and what we need to do it in the future.}
{The method is based on a simple mathematical model of a super-hot  ($T\ga 10^8$~K)
turbulent-current layer (SHTCL) and  a model of plasma heating by the layer.}
{The models allow us to study a correspondence between the main
characteristics of the layer, such as temperature and
dimensions, and the observational features, such as differential and
integral emission measure of heated plasma, intensity of spectral
lines \ion{Fe}{xxvi} (1.78 and 1.51~\AA) and \ion{Ni}{xxvii}
(1.59~\AA). This method provides a theoretical basis for
determining parameters of the current layer from observations.}
{Observations of SHTCLs are difficult, because the spectral
line intensities are faint, but it is theoretically possible in
the future. Observations in X-ray range 1.5--1.8~\AA ~with high
spectral resolution (better than 0.01~\AA) and high temporal
resolution (seconds) are needed. It is also very important to
interpret the observations using a multi-temperature approach
instead of the usual single or double temperature method.}


\titlerunning{Current-layer diagnostics in solar flares}

\maketitle


\section{Introduction}

Solar flares are the brightest events of solar activity. High
energy, about $2\times 10^{32}$~ergs, is released
during a typical flare;  it is expended on heat fluxes, particle
acceleration, plasma flows, etc. The basis of this phenomenon is
magnetic reconnection, which converts the magnetic-field energy
into other forms. The conversion occurs in a reconnecting current
layer. It is formed in the place where opposite-directed
magnetic fluxes of an active region interact among themselves
(e.g., Priest \& Forbes 2000; Somov 2006b).

Rich information about solar flares can be obtained from
observations in X-rays, which arise due to radiation of
high-temperature ($10^7 - 10^8$~K) plasma and accelerated
(up to MeV energies) electrons.
Spacecrafts, such as Hinotori, Yohkoh, Solar Maximum Mission (SMM),
and Reuven Ramaty High Energy Solar Spectroscopic Imager (RHESSI),
provide indirect evidence of magnetic reconnection in flares:
cusp-shaped coronal structures,
shrinkage of newly reconnected field lines,
reconnection inflows and outflows,
a hard X-ray source above the soft X-ray loop top,  and others
(Emslie et al. 2004; Fletcher et al. 2011;
 Hudson 2011 and references therein).

Sui \& Holman (2003) describe the limb flare on 2002 April 15.
The double-source structure was observed  by RHESSI
in the range of 10--20 keV.
The temperature of the lower source, corresponding to flare loops,
increased toward higher altitudes, while the temperature of the
coronal source increased toward lower altitudes.
This is interpreted as a current layer formation between the two
sources.
We note that the central part of the structure, the super-hot
turbulent-current layer (SHTCL), remains empty on the images.
A similar pattern was also observed in the flare on 2002 April 30
(Liu et al. 2008).
Caspi \& Lin (2010) report RHESSI observation of the flare
on 2002 July 23.
A super-hot elongated source at $\sim 4\times 10^7$~K
is situated higher and separately from a colder source
at $\sim 2\times 10^7$~K.
It also agrees well with a concept of a current layer located
higher in the corona.

The first image of a large-scale reconnecting current layer
is presented by
Liu et al. (2010).
They describe a bright  structure of length $>0.25R_\odot$ and
width $(5-10)\times 10^8$~cm, which is situated above the cusp-shaped
loop at heights $\sim (0.78-1.05)R_\odot$ and associated with a
coronal mass ejection (CME).
It was observed by extreme ultraviolet imaging telescope (EIT)
onboard Solar and Heliospheric Observatory (SOHO).
Its 195~\AA  ~channel is sensitive to low-temperature
($1.6\times 10^6$~K) and to high-temperature
($2\times 10^7$~K) plasma because of the presence of the
\ion{Fe}{xii} (195~\AA) and \ion{Fe}{xxiv} (192~\AA)
resonance lines.
The super-hot ($\ga 10^8$~K) plasma cannot be registered using EIT.
The characteristic density of the layer is
$\sim 10^6-10^7$~cm$^{-3}$.
In our work, we consider more compact current layers
that are associated with flares and that are formed in the lower corona
at heights  $\sim 0.05\, R_\odot$ in strong magnetic fields.

Models of the current layers are developed analytically (Sweet
1958; Parker 1963; Syrovatskii 1963, 1966; Oreshina \& Somov 2000)
and numerically (Chen et al. 1999; Yokoyama \& Shibata
2001; Reeves et al. 2010; Shen et al. 2011). The advantage of
numerical computations consists in the possibility of considering
various effects: heat conduction, plasma motion, radiative
losses etc. It allows one to clarify qualitative role of the
effects but  does not give real quantitative estimations. Modern
numerical MHD algorithms cannot work with real transport
coefficients replacing them with some fictive values. They always
contain an artificial numerical viscosity and conductivity. As a
result, computations of fine structure of the layer, whose
thickness is less than its width and length by 7 orders of
magnitude, become impossible. Analytical models are free from
the numerical drawbacks. As they cannot simultaneously consider
many effects, it is very important to determine the main
one. Their advantage consists in simplicity and clarity.
Analytical models also help to interpret the results of more
complicated numerical modelling. They can serve as a test for these
computations. Without doubt, both numerical and analytical
approaches have to work together, giving useful ideas to each
other.

As we have already noticed, we consider SHTCLs in large solar
flares. These layers are expected to be significantly hotter,
denser, and more compact than those associated with CME, because
they are situated at lower altitudes in the corona. According to
the analytical self-consistent model of a SHTCL, their temperatures are
$ \ga 10^8$~K, and their thicknesses are  $\la 1$~m (Somov 2006b). Their
emission measures are too low for direct observations. Meanwhile,
these observations are important, because spatial and temporal
averaging can lead to ambiguity and erroneous
interpretation. For example, Doschek \& Feldman (1987) demonstrate
various plasma temperature structures leading to the same X-ray
spectra; Li \& Gan (2011) have shown that the summation of two
single-power X-ray continuum spectra from different sources looks
like a broken-up spectrum. The direct observations are necessary
for understanding the basic physics and refinement theoretical
models.

Our study is based on simple models of the SHTCL and
plasma heating by heat fluxes from it.
We note that heat conduction is not the only mechanism of plasma
heating.
Electrons, accelerated in a current layer, can also make
significant contribution.
The relative role of each mechanism is different in different
flares (Veronig et al. 2005).
Kobayashi et al. (2006) report about purely thermal flare
on 2002 May 24  that was observed in the 20--120~keV range by ballon-born
hard X-ray (HXR) spectrometer.
Moreover, heat conduction seems to be dominant at pre-impulsive
phase; namely, it is a gradual rise phase that precedes the impulsive phase
(Farnik \& Savy 1998;
Veronig et al. 2002;
Battaglia et al. 2009).

Thus, our study is applicable primarily for the beginning of the flare.
This phase is of particular interest due to the absence of
accompanying processes like chromosphere evaporation and others
which can affect the reconnection region.
They significantly complicate the overall picture, interpretation
of observations, and theoretical models
(for example, Liu et al. 2009).
We would like to see 'pure' current layers and their vicinities.
The aim of our work is to clarify why we actually do not
directly observe current layers in large solar flares
and what we need to do it in the future.

As shown further, we need to consider {\it
multitemperature and nonequilibrium effects in plasma}. They are
small; it is difficult to observe them. Often, it is difficult
even to distinguish them from instrumental effects (e.g. Holman et
al. 2011, Kontar et al. 2011). Perhaps, this is one reason why many
authors study mainly mean values of flare plasma. For example,
Phillips et al. (2006) investigated spectra obtained mostly during
the flare decay phase ``to minimize instrumental problems with
high count rates and effects associated with multitemperature and
nonthermal spectral components''. Meanwhile, the authors note
that the isothermal model gives poor fits for flares near their
maximum and propose a possible explanation, which consists in
departing the emitting plasma from the isothermal state. Thus,
sometimes we can see evidence of the multitemperature
nonequilibrium plasma. These events must be analyzed very
accurately.

The paper is organized as follows.
In Sect.~2, we describe an analytical model of the SHTCL
and a model of plasma heating by the layer.
In Sect.~3, observational manifestations of the SHTCL
(emission measure of heated plasma,
intensities of some spectral lines, etc.) are presented.
In the last section, we formulate our conclusions.


\section{Models of current layer and plasma heating}

\subsection{Current layer}

Our study is based on the simple analytical model
of the super-hot turbulent-current layer (Oreshina \& Somov 2000)
with the heat flux by Manheimer \& Klein (1975).
They investigated the heat transport in a plasma heated impulsively by a
laser and obtained the following expression for the heat flux limitation:
\[
F = C_F \, n_{\rm e}\, T_{\rm e}^{3/2}\, ,
\]
where
\begin{equation}
C_F = (0.22 - 0.78)\times 10^{-10}\,\, \mbox{erg\, cm\, s}^{-1}\,
\mbox{K}^{-3/2} . \label{eq-CF}
\end{equation}
In this work, we take
$C_F = 0.22\times 10^{-10}\,\, \mbox{erg\, cm\, s}^{-1}\, \mbox{K}^{-3/2}$.

Magnetic flux tubes  with plasma move (from above and from below)
at low velocity $v$ toward the SHTCL, penetrate into it, reconnect at the
layer center, and then move along the $x$-axis toward its edges,
accelerating up to high velocity  $V$
(Fig.~1).
Our model does not consider the internal structure of the layer.

\begin{figure}
 \centerline{\includegraphics[width=88mm]{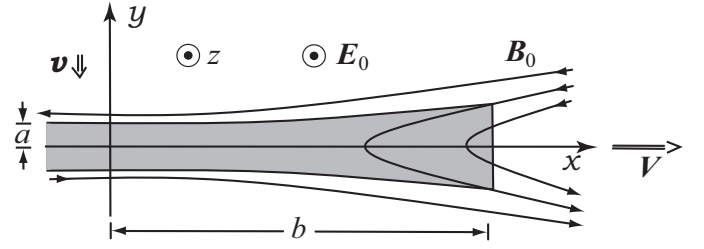}}
 \caption{Reconnecting current layer and field lines.}
\end{figure}

The input parameters are
electron number density $n_0$ outside the layer,
the gradient of the magnetic field $h_0$ in the vicinity of separator,
the inductive electric field $E_0$,
and the ratio $\xi=B_y/B_x$ of the transverse and main components
of magnetic field.
The mass, momentum, and energy conservation lows give
the following analytical expressions for the layer parameters:

\noindent
for the half-thickness,
\[
a = 1.7\times 10^6 \times n_0^{-1/2}, \,\, \mathrm{cm;}
\]
for the half-width,
\[
b = 0.22 \times n_0^{1/4}\, h_0^{-1} \, E_0^{1/2}\, \xi^{\, -1/2},\,\, \mathrm{cm;}
\]
for the electron temperature,
\[
T_{\rm s} = 1.8\times 10^{12} \times n_0^{-1/2} \, E_0\, \xi^{\, -1}, \,\, \mathrm{K;}
\]
for the electron density,
\[
n_{\rm s} = 7.9 \times n_0, \,\, \mathrm{cm}^{-3};
\]
for the magnetic field on the inflow sides of the current layer,
\[
B_0 = 0.22 \times n_0^{1/4} \, E_0^{1/2} \, \xi^{\, -1/2}, \,\, \mathrm{G};
\]
for the inflow velocity,
\[
v = 1.4\times 10^{11} \times n_0^{-1/4} \, E_0^{1/2} \, \xi^{\, 1/2},\,\, \mathrm{cm\, s}^{-1};
\]
for the outflow velocity,
\[
V = 1.7\times 10^{10} \times n_0^{-1/4} \, E_0^{1/2}\, \xi^{\, -1/2},\,\, \mathrm{cm\, s}^{-1};
\]
and for the heating time, which is the time for a given magnetic tube to be connected to the SHTCL,
\[
t_{\mathrm h} = \frac{2b}{V} = 2.6 \times 10^{-11} \times n_0^{1/2}\, h_0^{-1} \, , \,\, s.
\]

\noindent
Table~1 presents two sets of input parameters, which are specific to solar flares,
and corresponding output values.
We consider two input sets to emphasize the diagnostic
capabilities of our models.
In the Set~2, only the electric field
is decreased when compared with the Set~1.
This leads to decreasing the temperature $T_{\mathrm s}$, while
the heating time $t_{\mathrm h}$ is not changed.
Both these parameters are important for our further computations of
plasma heating by the SHTCL.

The input and output parameters agree with the following
observational and theoretical results. The electron number density
is consistent with statistical data from the overview of flares by
(Caspi 2010): $(0.8 - 30)\times 10^{10}~\mathrm{cm}^{-3}$.
Fig.~5.5 of that work (page 70, right panel) presents densities of
super-hot plasma versus maximum RHESSI temperatures that can be
as high as 45-50 MK. We note that they are estimated using an
isothermal approach. The multitemperature method gives higher
temperatures of some portion of plasma.

The electric field agrees with estimations by Somov et al. (2008),
Lin (2011): $E_0 \sim 10 - 30~\mathrm{V\,cm}^{-1} \approx 0.03 -
0.1~\mathrm{CGSE}$. The ratio $\xi$ belongs to the range of values
where the current layer is stable (Somov \& Verneta 1993).

Magnetic fields up to 350~G are expected in coronal sources (Caspi
2010, Fletcher et al. 2011).
We assume that these fields are in the
magnetic-reconnection region, which is in the vicinity of a separator
at heights up to $\sim 4\times 10^9$ cm.
At present, there are no reliable direct measurements of the coronal
magnetic fields.
Our theoretical estimations
show that magnetic fields can achieve even higher values, up to
900~G, at the separator (Oreshina \& Somov, 2009).

We note that the thickness of the layer is very small,
and hence, magnetohydrodynamics (MHD) is not a good approximation.
We should solve the kinetic equation,
but this significantly complicates the model.
Instead, we use anomalous transport coefficients,
thus considering the deviation from the MHD.
Relaxation heat flux (Eq.~\ref{eq-F-relax}) is also due
to the deviation from the Maxwell distribution function.


\begin{table}
\caption{Two sets of parameters for the current layer model}
\begin{tabular}{lcc}
\hline\hline
                        & Set 1             & Set 2             \\
\hline
Input parameters        &                   &                   \\
$n_0, \,\,\mathrm{cm}^{-3}$    & $10^{10}$         & $10^{10}$  \\
$h_0, \,\,\mathrm{G\,\, cm^{-1}}$ & $1.2\times 10^{-6}$ & $1.2\times 10^{-6}$ \\
$E_0, \,\,\mathrm{CGSE}$& $0.1$             &  $0.0425$            \\
$\xi = B_y/B_x$         & $0.005$          &  $0.005$           \\
\hline
Output parameters           &                   &                    \\
$a, \,\,\mathrm{cm}$           & $17$              & $17$               \\
$b, \,\,\mathrm{cm}$           & $2.6\times 10^8$  & $1.7\times 10^8$   \\
$T_{\mathrm s}, \,\,\mathrm{K}$   & $3.5\times 10^8$  & $1.5\times 10^8$   \\
$n_{\mathrm s},\,\,\mathrm{cm}^{-3}$ & $7.9\times 10^{10}$  &  $7.9\times 10^{10}$ \\
$B_0, \,\, \mathrm{G}$         & $312$             & $204$              \\
$v, \,\,\mathrm{cm\, s}^{-1}$  & $9.6\times 10^6$  & $6.3\times 10^6$   \\
$V, \,\,\mathrm{cm\, s}^{-1}$  & $2.4\times 10^8$  & $1.6\times 10^8$   \\
$t_{\mathrm h}, \,\,\mathrm{s}$   & $2$               & $2$              \\
\hline
\end{tabular}
\end{table}


\subsection{Plasma heating}

Let us now consider heating the surrounding plasma by the SHTCL.
Due to conversion of magnetic-field energy, the layer temperature
is very high, of about $3\times 10^8~\mathrm{K}$
(Table~1). When a magnetic tube comes into contact with the layer,
plasma inside is heated. While the tube moves from the
center to the edge of the layer, a large amplitude heat wave
travels along it. After the disconnection from the heating
source, the tube continues to move in the corona. Now, only the
energy redistribution occurs inside it.

The properties of these thermal waves are complex because of
hydrodynamic flows of emitting plasma and kinetic phenomena in it
(Somov 1992, Ch.~2).
Here, we consider the simplified model to emphasize the main
effect, heat conduction.
This process can be described by a simple equation
(Somov 1992, p.~169):
\begin{equation}
    \frac{ \partial \, \varepsilon }{ \partial \, t} \, = - \,
    \mathrm{div} \: {\vec F} ,
    \label{eq-conduction}
\end{equation}
where $\vec F$ is the thermal energy flow density,  and
$\varepsilon$ is the internal energy of unit plasma volume,
which is the energy of chaotic motion of particles.
We consider only the electron heating
because the timescale of the thermal wave propagation
is too short to heat ions.
Indeed, the time of collisions for electrons is
(e.g. Somov 2006a, Sect. 8.31, p.~142)
\[
\tau_{\rm ee} = \frac
{m_{\rm e}^2\,(3\, k_{_{\rm B}} T_{\rm e}/m_{\rm e})^{3/2}}
{\pi\, e_{\rm e}^4 \, n_{\rm e}\, (8\, {\rm ln}\Lambda)} \,
\frac{1}{0.714} \, ,
\]
where $m_{\rm e}$, $e_{\rm e}$, $n_{\rm e}$, and $T_{\rm e}$ are
the respective electron mass, charge, density, and temperature; $k_{_{\rm
B}}$ is the Boltzmann constant; and ${\rm ln}\Lambda = {\rm ln}\left (
9.44 \times 10^{\, 6} \, T_{\rm e}   / n_{\rm e}^{\,\, 1/2}\right
)$ is the Coulomb logarithm for $T_{\rm e} > 5.8\times 10^5$~K.
The ion collision time is
\[
\tau_{\rm ii} = \frac
{m_{\rm i}^2\,(3\, k_{_{\rm B}} T_{\rm i}/m_{\rm i})^{3/2}}
{\pi\, e_{\rm i}^4 \, n_{\rm i}\, (8\, {\rm ln}\Lambda)} \,
\frac{1}{0.714} \, ,
\]
where $m_{\rm i}$, $e_{\rm i}$, $n_{\rm i}$, and $T_{\rm i}$ are
the respective ion mass, charge, density, and temperature.
The characteristic time of temperature equalizing between the electron
and ion components in plasma is
\[
\tau_{\rm ei} = \frac
{m_{\rm e}\, m_{\rm i}\,(3\, k_{_{\rm B}} (T_{\rm e}/m_{\rm e}+T_{\rm i}/m_{\rm i})^{3/2}}
{(6\, \pi)^{1/2}\, e_{\rm e}^2\, e_{\rm i}^2 \, n_{\rm i}\, (8\, {\rm ln}\Lambda)}  \, .
\]
Therefore, we get
$\tau_{\rm ee}\sim 1$~s,
$\tau_{\rm ii}\sim 50$~s, and
$\tau_{\rm ei}\sim 1100$~s
for the electron and proton components of plasma with temperature
$\sim 10^8$~K and density $\sim 10^{10}$~cm$^{-3}$.
Thus, Coulomb collisions between electrons
in the vicinity of the SHTCL
are much more frequent than between protons;
energy exchange between electrons and protons is very slow.
The thermodynamic equilibrium is achieved first for electrons, then for protons, and
finally for the whole plasma.

Thus, the internal energy in Eq.~(\ref{eq-conduction}) is
\begin{equation}
 \varepsilon = \frac{3}{2}  \, n_{\mathrm e} k_{_{\mathrm B}} T \, ,
 \label{eq-int_energy}
\end{equation}
where $T$ is the electron temperature.
This state may not be reached in the conditions under consideration.
We use Eq.~\ref{eq-int_energy} to define the
effective temperature.
In the frame of kinetic theory, it is a measure of the width
of the electron distribution function in velocity space
(Somov 2006a, p.~171).

Various types of heat flux in the SHTCL vicinity have been
considered in detail by Oreshina \&  Somov (2011). The classical
Fourier expression ${\vec F}=- \kappa \nabla T$ is not valid here,
because it is derived for plasma in a state close to local
thermodynamic equilibrium. The latter suggests that the
characteristic timescale of the process is much longer than the
electron collision time $\tau_{\rm ee}$ and the characteristic
length scale exceeds greatly the electron mean free path
$\lambda_{\rm ee}$. These conditions are violated at the
temperature of about $10^8$~K. The thermal wave is too fast and
has too steep front, so that the characteristic time of the
wave propagation $t_T=\frac{T}{\partial T / \partial t}$ is less
than $\tau_{\rm ee}$, and the characteristic length scale
$l_T=\frac{T}{\partial T /\partial l}$ is less than $\lambda_{\rm
ee}$ by $1-3$ orders of magnitude. The corresponding plots are
presented in the paper by Oreshina \&  Somov (2011).

We also have showed that there is a generalization of the
classical Fourier law, which is better suited to solar flares. It
is described by Moses \&  Duderstadt (1977), who applied it to a
laser heated plasma. A general consideration of the problem can be
found in (Shkarovskii et al. 1969; Golant et al. 1977). The method
is developed in a fundamental framework and is physically
substantiated. It is not restricted to slow time variations in the
distribution function and thus better suited to the treatment of
rapidly varying thermal processes.
The technique is based on the Grad 13 moment equations.
The electron distribution function is expanded in terms of
Hermite-Chebyshev polynomials.
The thirteen moments of the distribution function have a clear physical
meaning -- namely, the density, velocity, temperature, pressure tensor,
and heat flux.
The system of equations for them is derived from the Boltzmann equation and
includes the mass, momentum, and energy conservation laws,
the equation for the pressure tensor, and,
finally, the equation for heat flux:
\begin{equation}
    \vec{F} = - \kappa \, \nabla T \, -
    \, \tau \, \frac{ \partial \vec{F} }{ \partial t } \, .
    \label{eq-F-relax}
\end{equation}
Here,
$\kappa$ is the heat conductivity coefficient,
$ \tau $ is some characteristic relaxation time.

According to
Braginskii (1963, v.~1, p.~183),
\begin{equation}
\kappa \, =  \,  \kappa_0 \, T^{5/2} \, ,  \qquad
\tau \, = \, \tau_0 \,  \, T^{3/2}\, ,
\label{eq-kappa-cl}
\end{equation}
where
$
   \kappa_0 = 1.84 \times 10^{\, -5} /\,  {\mbox{ln} \, \Lambda}
   \, .
$

The technique for calculating the constant $\tau_0$ is
described by Oreshina \& Somov (2011).
\[
\tau_0 = \frac{2}{3}\frac{k_{_{\rm B}}\, \kappa_0}{ n_{\rm e} \,
C_F^{\, 2}} \, ,
\]

\noindent where $C_F$ is determined by equality (\ref{eq-CF}).
For the solar flare conditions, we estimate
\[
\tau_0 \approx ( 0.11 - 1.3) \times 10^{-11} \,\,
   \mathrm{s \, K}^{-3/2}  .
\]
The corresponding relaxation time at the temperature of about $10^8$~K is
\[
   \tau = \tau_0 \, T^{3/2}\big|_{T\sim 10^8} \sim 1-10 \,\, \mathrm{s} \, .
   \label{tau}
\]
These values are comparable to the time of tube contact with the SHTCL,
$t_{\mathrm h} \approx 2$~s (Table~1).
Therefore, the collisional relaxation of heat flux must be considered
in models of plasma heating in solar flares.
In the following computations, we use
$\tau_0 = 1.3 \times 10^{-11}~\mathrm{s\, K^{-3/2}}$.

Thus, we solve both the heat conduction equation (\ref{eq-conduction})
and the equation for heat flux (\ref{eq-F-relax}),
which determine the temperature $T(\vec{r},t)$ and the
heat flux $\vec{F}(\vec{r},t)$.
For simplicity, we take
$n_{\mathrm e} = \mathrm{const} = 10^{10}~\mathrm{cm}^{-3}$.
We also assume that magnetic tubes
are straight and have a constant cross-sectional area.
Equations~(\ref{eq-conduction}) and  (\ref{eq-F-relax})
are rewritten in one-dimensional form:
\begin{equation}
     \frac{ 3 }{ 2 } \, n_{\rm e} k_{_{\rm B}}
     \frac{ \partial \, T }{ \partial \, t } \,
 = - \,\frac{ \partial \, F }{ \partial \, l } \, ,
   \label{eq-conduction-1dim}
\end{equation}
\begin{equation}
    F = - \,\kappa_0\, T^{5/2} \, \frac{ \partial T }{ \partial l } \, -
    \, \tau_0\, T^{3/2} \, \frac{ \partial F }{ \partial t } \, ,
    \label{eq-F-1dim}
\end{equation}
where $ l $ is a coordinate measured from the layer along the tube.
We note that the SHTCL is very thin: the ratio of its thickness
to the width is less $3\times 10^{-8}$ (Table~1).
Thus, we can neglect its thickness and solve
Eqs.~(\ref{eq-conduction-1dim}) and  (\ref{eq-F-1dim})
for the following boundary conditions.

1) For the time interval $0\lid t \lid t_\mathrm{h}$,
one end of the tube is connected to the SHTCL
with a temperature $T_\mathrm{s}$,
and the other end is far away and is maintained
at a coronal temperature  $T_0=10^6$~K:
\begin{equation}
T (l, t) \, \big| _{\, l = 0} = T_{\rm s} \, , \quad
    T (l, t) \, \big| _{\, l \rightarrow \infty } = T_0 \, .
     \label{eq-gr1}
\end{equation}

2) For $t > t_\mathrm{h}$, the tube is disconnected from the layer and
does not receive any heat from it:
\begin{equation}
    {F(l,t) \, \big| } _{\, l = 0} = 0 \, , \quad
    T(l,\, t) \, \big| _{\, l \rightarrow \infty } = T_0 \, ,
\end{equation}
\begin{equation}
    Q
    = \int \limits_{0}^{\infty } T(l,\, t) \,
    {\rm d} l = {\rm const} \, .
    \label{eq-gr2}
\end{equation}
Equation~(\ref{eq-gr2})
is the law of thermal-energy conservation in the tube after the
separation from the layer.

\begin{figure}
 \centerline{\includegraphics[width=88mm]{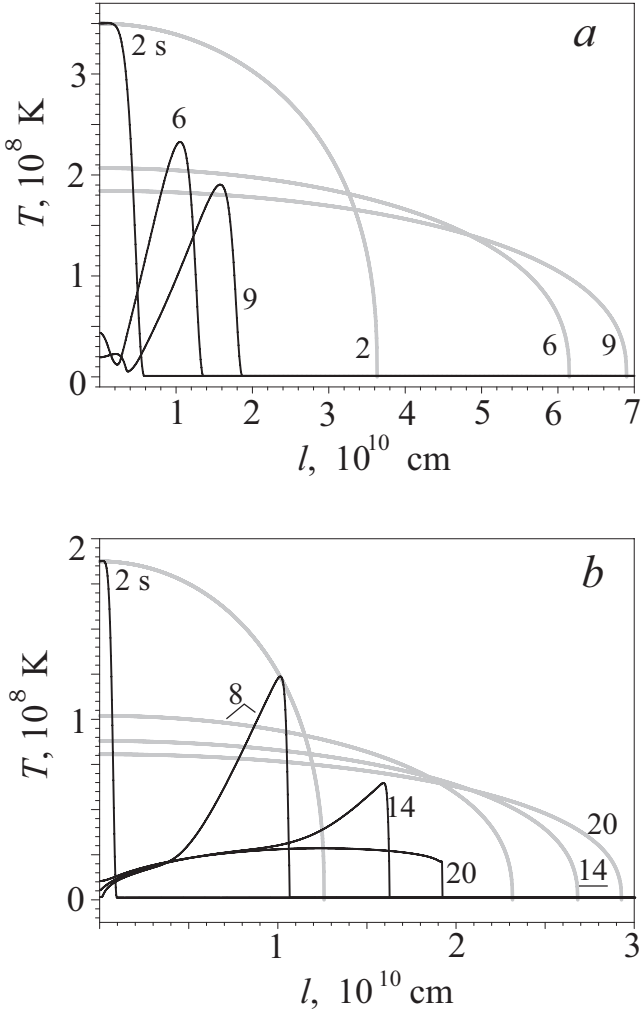}}
 \caption{Temperature distributions along a magnetic tube
  at consequent moments of time, which are obtained with (black curves)
  and without (grey curves) heat-flux relaxation.
  (a) $T_{\mathrm s} = 3.5\times 10^8$~K, (b) $T_{\mathrm s} = 1.5\times 10^8$~K.
  The numbers near the curves denote the time (seconds) passed after
  the beginning of the tube contact with the SHTCL.}
  \label{fig-temperature}
\end{figure}

The problem was solved numerically.
The obtained temperature distributions at consequent moments of time
are shown in Fig.~\ref{fig-temperature} by the black curves.
The grey curves indicate the corresponding distributions obtained
for the classical Fourier law, without the relaxation,
$\tau_0=0$.
The top panel presents the thermal wave for the first SHTCL model
(see Table~1) and the bottom panel for the second one.
They differ in the temperature of the current layer:
$T_{\mathrm s}=3.5\times 10^8$~K in the first case
and $T_{\mathrm s}=1.5\times 10^8$~K in the second one.
We stop computations when the heat wave passes
from the layer toward  the chromosphere at the distance  about
$1.9\times 10^{10}$~cm.
Thus, the computations were interrupted at $t=9$~s for the first model and
at $t=20$~s for the second one.
For more details, see Appendix.

Comparison between the relaxation and classical heat conduction
allows us to conclude the following.
The relaxation  limits the heat flux,
making it more realistic.
The amount of thermal energy $Q$, supplied to the tube
during its contact with the SHTCL, decreases by a factor of 10.
The thermal wave slows down as compared to the classical heat
conduction.
For example, Fig.~\ref{fig-temperature}(a) shows that the front velocity is about
$3\times 10^9~\mathrm{cm\,s}^{-1}$,
whereas it is  more than the speed of light
in the first second of tube contact with the SHTCL
in the classical case.
The wave shape is also changed after the tube disconnection
from the SHTCL.
In the relaxation case, the temperature maximum is at the wave front
and moves along the tube; a steep front is followed by a smooth decrease.
In the classical case, the temperature is at a maximum at the
point $l=0$.
The relaxation effect decreases in time and the wave shape becomes
almost classical (Fig.~\ref{fig-temperature}(b)).

Comparison between the cases with different temperature of the layer
shows that the amount of heat $Q$ in the tube is less
by a factor of 2 in the second model.
This result is expected because the temperature of the layer in the second model
is $1.5\times 10^8$~K  instead of $3.5\times 10^8$~K  in the first model.
We note that the temperature behind the wave  in the first seconds
is close to the background, and we have a solitary wave.
In course of time, the temperature behind the wave is increased, and
the wave becomes similar to the classic wave
about 20 seconds after the disconnection from the layer
(Fig.~\ref{fig-temperature}(b)).
The role of the relaxation is decreased in time.


\section{Observational manifestations of the current layer}
\subsection{Emission measure of the heated plasma}

\begin{figure}
 \centerline{\includegraphics[width=88mm]{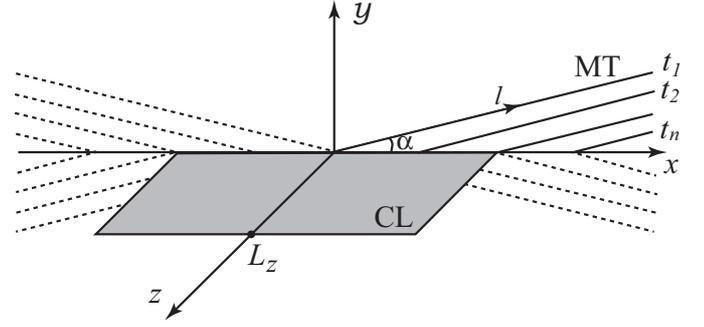}}
 \caption{Schematic representation of the radiative source.
 CL -- current layer, MT -- magnetic tubes,
 $t_1$, $t_2$ ... $t_n$ -- time passed from the tube penetration into the CL,
 $l$ - coordinate along a tube,
 $\alpha$ -- tube inclination to the CL plane.}
 \label{fig-radiative-source}
\end{figure}

The set of all the magnetic tubes, heated by the SHTCL, is a
source of radiation. To compute the emission measure of this
heated plasma, we consider a quarter of the tubes, shown by bold
lines in Fig.~\ref{fig-radiative-source}. The result is also
applicable to the other three quarters if we assume the process to
be symmetrical. Let all the tubes be in the $(x,\, y)$-plane and
parallel planes within the layer length $L_z$. For simplicity,
only the tubes in the $(x,\, y)$-plane are shown in
Fig.~\ref{fig-radiative-source}. In the layer vicinity, all the
tubes are modeled by straight lines, which are inclined at the same
angle $\alpha$ to the $(x,\, z)$-plane. In the frames of this
model, the magnetic tubes differ from each other only by the time
($t_1$, $t_2$ ... $t_n$) passed from their penetrating into the
layer.

By definition, the emission measure of the radiative source is
\begin{equation}
EM = \int\limits_\mathcal{V} n_\mathrm{e}^2 \, \mathrm{d} \mathcal{V} \, ,
\label{eq-EM-definition}
\end{equation}
where $\mathcal V$ is the volume of the heated plasma.
Assuming the tubes are distributed homogeneously
along the $z$-axis,
Eq.~(\ref{eq-EM-definition}) can be rewritten for
$x$, $y$, and $z$ coordinates:
\begin{equation}
EM  =
\int\limits_x \int\limits_y \int\limits_z n_\mathrm{e}^2 \,
\mathrm{d} x \, \mathrm{d} y \, \mathrm{d} z \, = \,
L_z\,   \int\limits_x \int\limits_y n_\mathrm{e}^2 \,
\mathrm{d} x \, \mathrm{d} y \, .
\label{eq-EM-xy}
\end{equation}
The length $L_z$ of the SHTCL is a new input parameter.
In the following computations, we use $L_z = 10^9$~cm.
We replace coordinates $x$ and $y$ by $t$ and $l$;
$t$ defines a tube and
$l$ is the point on it.
These are related by the equalities
\begin{eqnarray}
x & = & V\, t + l\, \mathrm{cos}\, \alpha \, , \\
y & = & l \,\, \mathrm{sin}\, \alpha \, ,
\end{eqnarray}
where $V$ is the velocity of tube movement along
the $x$-axis.
The Jacobian of this transformation is
\[
\left |
\begin{array}{cc}
\displaystyle \frac{\partial x}{\partial t} &
\displaystyle \frac{\partial x}{\partial l} \\[4mm]
\displaystyle \frac{\partial y}{\partial t} &
\displaystyle \frac{\partial y}{\partial l}
\end{array}
\right | =
\left |
\begin{array}{cc}
 \vphantom{\displaystyle \frac{\partial x}{\partial t}}
 V & \mathrm{cos}\,\alpha  \\[4mm]
 \vphantom{\displaystyle \frac{\partial x}{\partial t}}
 0 & \mathrm{sin}\,\alpha
\end{array}
\right | =
V\, \mathrm{sin}\,\alpha \, ,
\]

\noindent
and the expression~(\ref{eq-EM-xy}) for the emission measure
takes the form:
\begin{eqnarray}
EM & = &
L_z\,
\int\limits_0^t \int\limits_0^{l_\mathrm{f}}
n_\mathrm{e}^2 \,\, (V\, \mathrm{sin}\,\alpha )\,
\mathrm{d} l \, \mathrm{d} t \,  = \nonumber \\
 & = & L_z\, V\, \mathrm{sin}\,\alpha \,
\int\limits_0^t \int\limits_0^{l_\mathrm{f}}
n_\mathrm{e}^2 \,\,  \mathrm{d} l \, \mathrm{d} t\, .
\nonumber
\end{eqnarray}
Here, $l_\mathrm{f}$ is a coordinate of the thermal
wave front.
We note that we have a unique relationship between
coordinate $l$ and temperature  $T(t, l)$
for every fixed time $t$ (see Sec.~2).
Hence,
\begin{eqnarray}
EM & = &
L_z V\, \mathrm{sin}\,\alpha \,
\int\limits_0^t
\left( \int\limits_{T(t,0)}^{T(t,l_\mathrm{f})}
n_\mathrm{e}^2 \,\,
\frac{1}{\partial T(t, l) / \partial l}\,
\mathrm{d} T
\right ) \mathrm{d} t = \nonumber\\
 & = &
L_z V\, \mathrm{sin}\,\alpha \,
\int\limits_0^t
\left( \int\limits_{T(t,0)}^{T(t,l_\mathrm{f})}
\mathrm{dem}(t,T) \, \mathrm{d} T \right ) \mathrm{d} t .
\end{eqnarray}

\begin{figure}
 \centerline{\includegraphics[width=88mm]{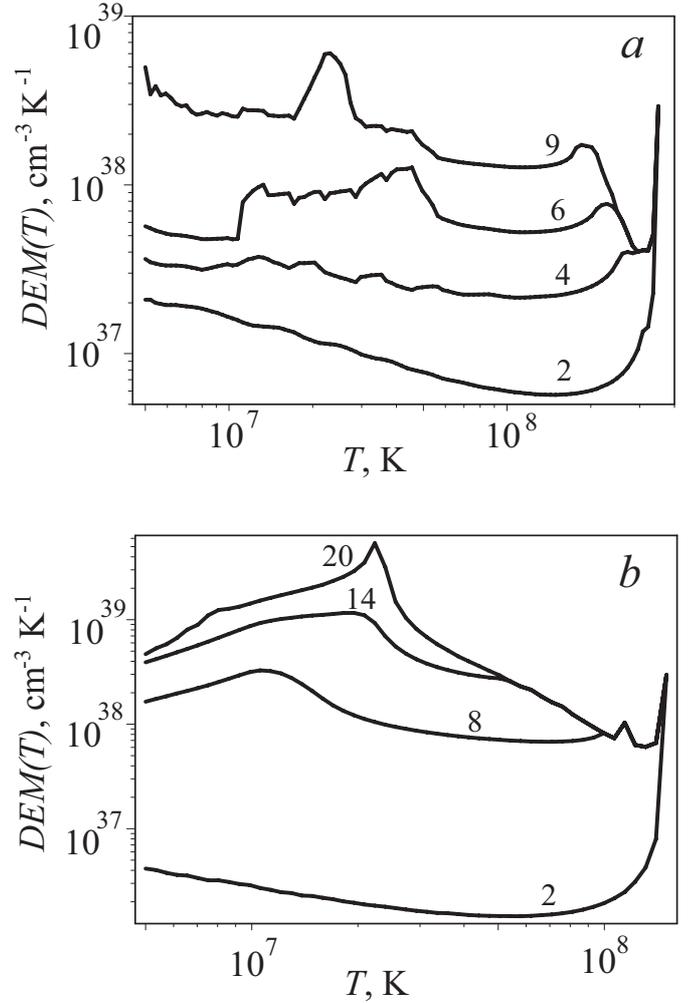}}
 \caption{Differential emission measure of the plasma heated
 by the SHTCL (a) for model 1 and (b) for model 2 (see Table 1).
 The numbers near the curves denote time (seconds)
 after the flare start.}
 \label{fig-DEM}
\end{figure}

\noindent
Function
\[
\left.
\mathrm{dem}(t,T)\equiv n_\mathrm{e}^2 \,\,
\frac{1}{\partial T(t, l) / \partial l}
\right|_{l=l(t,T)}
\]
is the differential emission measure of a single magnetic tube
of unit cross-section.
It is determined from the temperature distributions obtained
in Sec.~2.
The differential emission measure of the heated plasma is
\begin{equation}
DEM(t,T)  =  \frac{\mathrm{d}\, EM}{\mathrm{d}\, T} =
4\, L_z V\, \mathrm{sin}\,\alpha \,
\int\limits_0^t
\mathrm{dem}(t,T) \,
\mathrm{d} t.
\end{equation}
Here, the multiplier 4 considers the entire volume
of the heated plasma.
The differential emission measure for models 1 and 2
(Table~1) are presented in Fig.~\ref{fig-DEM} (a) and (b).

The behavior of the function $DEM(T)$  depends
on the parameters of the layer and hence can serve as a
diagnostic tool for it.
In the first model, we see  two  local maxima.
The narrow maximum at the layer
temperature $T\approx 3.5\times 10^8$~K is due to the super-hot current
layer and exists for every moment.
The wide maximum at  $T \approx (1-3)\times10^7$~K is gradually formed
due to a smooth temperature drop behind the wave front after the tube
disconnection from the layer.
In the second model,
a remarkable maximum is formed at $T\approx (1-3)\times 10^7$~K, too.
We note that these temperature estimations are usually
obtained from observations of super-hot plasma in solar flares
(Caspi 2010).
The local narrow maximum at the layer temperature $2\times 10^8$~K is
almost invisible, which  makes
direct observations of current layers difficult.

Differential emission measure of high-temperature plasma in
the M3.5 class flare was determined by Battaglia \& Kontar (2012).
The authors used  the observations of the Solar Dynamic Observatory (SDO)
in 94, 171, 211, 335, and 193~\AA ~spectral channels.
Thus, they cannot see the super-hot plasma with $T\sim 10^8$~K.
They consider only the plasma along the line of sight:
$DEM = n_\mathrm{e}^2 \, \mathrm{d} l/\mathrm{d} T, \quad
\mathrm{cm^{-5}\, K^{-1}}$.
Nevertheless, the maximum of $DEM(T)$ around $(1-3)\times 10^7$~K seems
to be common for our both methods.

\begin{figure}
 \centerline{\includegraphics[width=88mm]{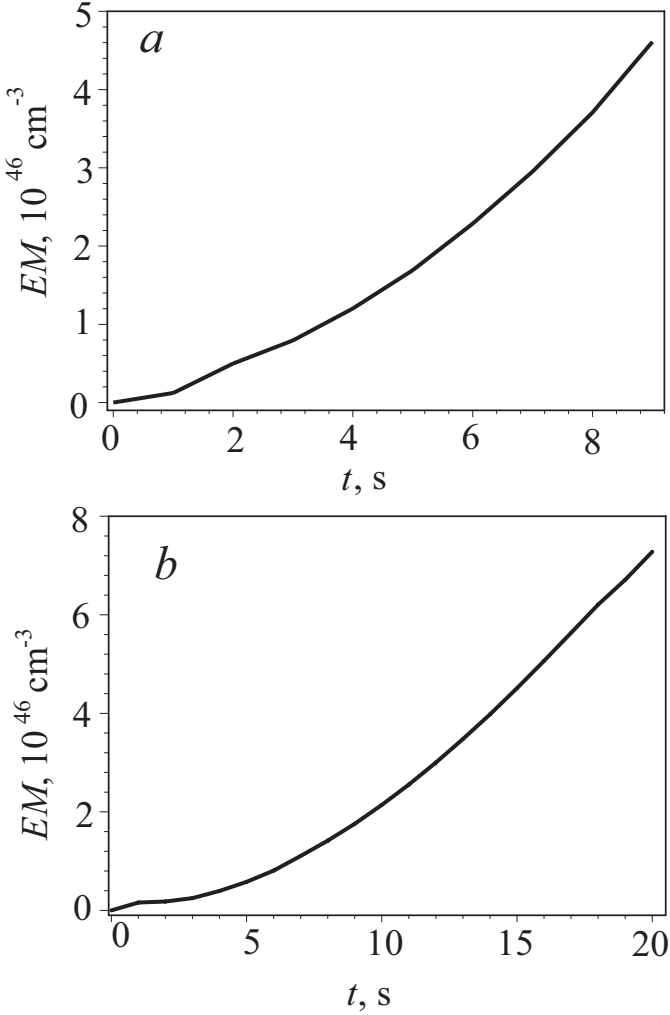}}
 \caption{Integral emission measure of the heated plasma
  (a) for model 1  and (b) for model 2. }
 \label{fig-EM}
\end{figure}

Fig.~\ref{fig-EM} presents the integral emission measure as a function of time:
\[
EM(t)=\int DEM(t,T)\,{\mathrm d}T\, .
\]
Our estimations are consistent by an order of magnitude
with RHESSI observations at the beginning of a flare
(Liu et al. 2008; Caspi \& Lin 2010).

\subsection{Spectral line intensities}

The obtained differential emission measure allows us to
compute intensities of spectral lines:
\begin{equation}
I_\lambda(t) = \frac{1}{4\, \pi \, R^2}
\int
\frac{P_\lambda (T)}{n_{\mathrm e}^2}\, DEM(t,T) \, \mathrm{d} T ,
\mathrm{photons \, cm^{-2}\, s^{-1}} .
\label{eq-intensity}
\end{equation}
Here,
$R$~(cm) is the distance from the Sun to the Earth, and
$P_\lambda(T)~(\mathrm{erg\, cm^{-3}\, s^{-1}})$ is  the line power, which is
the energy radiated by a unit volume of plasma per one second.

\begin{figure}
 \centerline{\includegraphics[width=88mm]{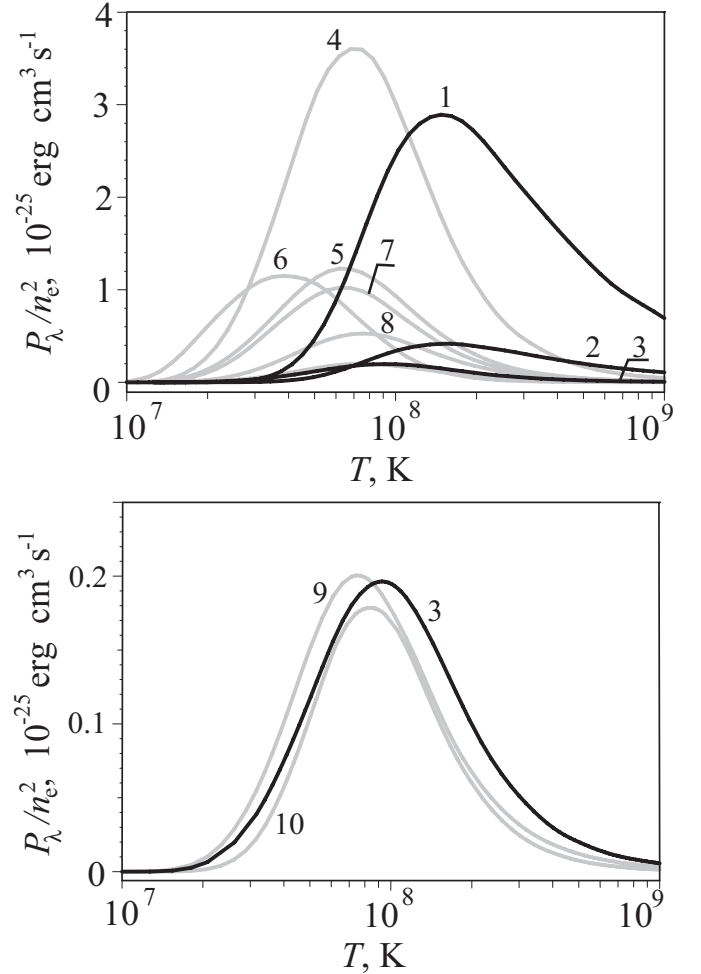}}
 \caption{Relative line power as a function of temperature,  drawn using the table
 by Mewe et al. 1985.
 The numbers near the curves denote the spectral lines according to Table~2.
 The black curves correspond to the super-hot spectral lines
 with $T_\mathrm{peak}\geq 10^8$~K and
 the grey curves to the lower-temperature ones with $T_\mathrm{peak}<10^8$~K.
 The weakest lines 3, 9, and 10 are also shown separately on the bottom panel.}
 \label{fig-line-power}
\end{figure}

The relative power $P_\lambda/n_{\mathrm e}^2$ of 2131 spectral lines in
X-rays (1.29--300 \AA) are computed
over the temperature range $(3\times 10^4) - 10^9$~K by
Mewe et al. (1985).
As we are interested in observing super-hot plasma,
we have selected lines with a power maximum at
$T \gid 10^8$~K.
The most powerful lines are
\ion{Fe}{xxvi} (1.78 \AA ~and 1.51 \AA) and \ion{Ni}{xxvii} (1.59 \AA).
They are listed in Table~2 (lines 1--3).
The classification of the transitions is given in usual form.
${(P_\lambda /n_{\mathrm e}^2)}_\mathrm{max}$ is the maximum power value, reached at the
temperature $T_\mathrm{peak}$.
Figure~\ref{fig-line-power} presents the relative power of the lines
as a function of temperature.
The super-hot lines 1--3 are shown by black curves.

\begin{table*}
\begin{minipage}{150mm}
\caption{Spectral lines for super-hot plasma observations.}
\label{table-lines}
\begin{tabular}{cccccc}
\hline\hline
No. & Ion & $\lambda$,~\AA & Transition
& ${(P_\lambda /n_e^2)}_\mathrm{max}$,&  $T_\mathrm{peak}$, \\
&     &   &          & $~ 10^{-25}\mathrm{erg\, cm^{3}\, s^{-1}}$ &  $10^8$~K  \\
\hline
1   & \ion{Fe}{xxvi} & 1.780 & $1s\, g2S1/2 - 2p\,\, 2P1/2,3/2$ & 2.9 & 1.58 \\
2   & \ion{Fe}{xxvi} & 1.510 & $1s\, g2S1/2 - 3p\,\, 2P1/2,3/2$ & 0.42 & 1.58 \\
3   & \ion{Ni}{xxvii}& 1.590 & $1s^2\, g1S0 - 1s2p\,\, 1P1$ & 0.19 & 1.00 \\
\hline
4   & \ion{Fe}{xxv}  & 1.851 & $1s^2\, g1S0 - 1s2p\,\, 1P1$ & 3.5  & 0.63 \\
5   & \ion{Fe}{xxv}  & 1.858 & $1s^2\, g1S0 - 1s2p\,\, 3P2,1$ & 1.23 & 0.63 \\
6   & \ion{Fe}{xxiv} & 1.864 & $1s^22p - 1s2p^2,\,\, 1s2s^2$ & 1.10 & 0.34 \\
7   & \ion{Fe}{xxv}  & 1.869 & $1s^2\, g1S0 - 1s2s\,\, 3S1$ & 1.00 & 0.63 \\
8   & \ion{Fe}{xxv}  & 1.570 & $1s^2\, g1S0 - 1s3p\,\, 1P1$ & 0.52 & 0.79 \\
9   & \ion{Fe}{xxv}  & 1.510 & $1s^2\, g1S0 - 1s4p\,\, 1P1$ & 0.20 & 0.79 \\
10  & \ion{Fe}{xxv}  & 1.790 & $1s2l - 2p2l$ ($l=s,p$)      & 0.18 & 0.79 \\
\hline
\end{tabular}
\end{minipage}
\end{table*}

There are many other lines in the spectral vicinity of the super-hot
lines 1--3, which are close in power but radiate at the lower temperatures,
$T_\mathrm{peak} < 10^8$~K
(lines 4--10 in Table~2, grey curves in Fig.~\ref{fig-line-power}).
For example, the difference between  the strongest lines of the two groups
is $\lambda_1 - \lambda_4 = 0.071~\mathrm{\AA}$.
Thus, it would be very valuable to have spectral resolution that allows one
to separate these lines from each other.
For fruitful SHTCL observations, the spectral resolution must be
better 0.01~\AA.
The spacecraft-borne instruments, such as X-ray crystal spectrometers onboard
the NASA Solar Maximum Mission and Japanese Hinotori spacecraft,
have even better resolution, allowing one to resolve Ly$_\alpha$ lines from
hydrogenic iron, $1s^2S_{1/2} - 2p^2P_{3/2}$ and $1s^2S_{1/2} - 2p^2P_{1/2}$
that occur at 1.778 and 1.783 \AA.

\begin{figure}
 \centerline{\includegraphics[width=88mm]{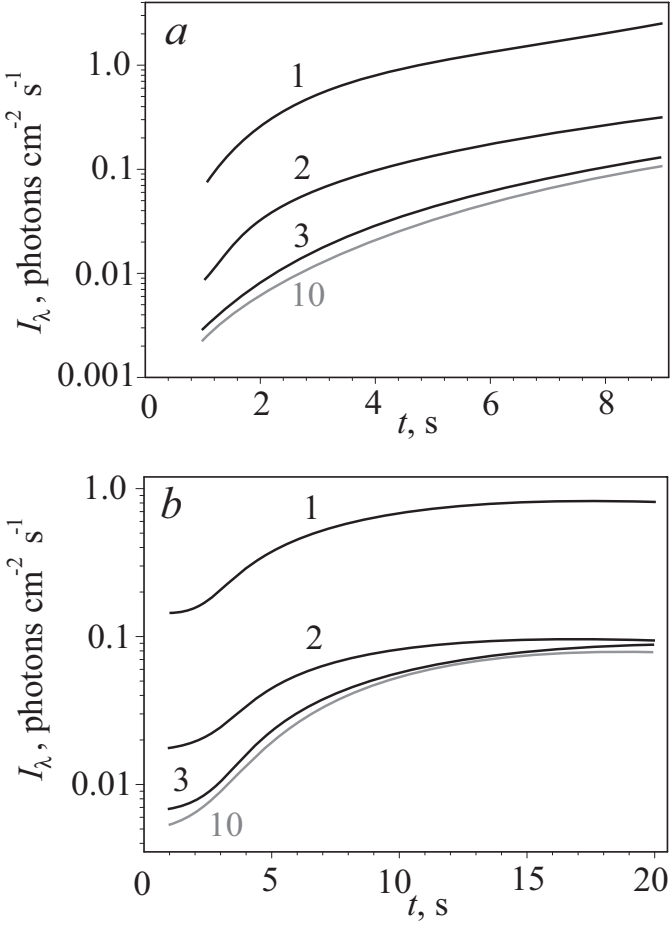}}
 \caption{Dependence of intensities of the super-hot spectral lines on time
 (a) for set~1 of SHTCL parameters  and  (b) for set~2.
 The numbers near the curves denote the spectral lines, according to Table~2.}
 \label{fig-intensity}
\end{figure}

Moreover, the sensitivity of the detectors should be sufficient
to record weak photon fluxes.
Figure~\ref{fig-intensity} shows the intensities of the super-hot lines
$1-3$ (black curves) computed using Eq.~(\ref{eq-intensity}).
They are faint.
Only the resonance line 1 has intensity
$I_\lambda \sim 0.1 - 2 ~\mathrm{photons}\, \mathrm{cm}^{-2}\,\mathrm{s}^{-1}$
whereas line 3, for example, is characterized by
$I_\lambda  \sim 0.01-0.1~\mathrm{photons}\,\, \mathrm{cm}^{-2}\,\mathrm{s}^{-1}$.

We note that the line power $P_\lambda$ was calculated by
Mewe et al. (1985)
under the assumption of ionization equilibrium;
that is
the ionization rate from the stage $j$ to $j+1$ is equal to the
recombination rate from  $j+1$ to  $j$:

\[
n_{\rm e} \, N_j\, Q_j =n_{\rm e} \, N_{j+1} \alpha_{j+1} \, ,
\]

\noindent
where $Q_j$ and $\alpha_j$ are the rate coefficients
of ionization and  recombination from the ionization stage $j$.
The ionization state $N_j$ of an ion does not depend on the ion
temperature $T_{\rm i}$,
but it is a function of the electron temperature $T_{\rm e}$.
We can estimate the characteristic times of ionization and recombination
using the formulas by Phillips (2004):
\[
\tau_{\rm ion} = \frac{1}{Q_j\, n_{\rm e}} \, , \qquad
\tau_{\rm rec} = \frac{1}{\alpha_j\, n_{\rm e}} \, .
\]
$Q_j$ and $\alpha_j$  were computed using the method by
Arnaud \& Raymond (1992).
We do not present here the formulas, because they are very cumbersome.
For ions \ion{Fe}{xxv} and \ion{Fe}{xxvi} at the temperature $T_{\rm e}\sim 10^8$~K
and density $n_{\rm e}\sim 10^{10}$~cm$^{-3}$,
we obtain $\tau_{\rm ion}\sim \tau_{\rm rec} \sim 100$~s.
Thus, the ionization equilibrium is not achieved during rapid
plasma heating in the vicinity of SHTCL.
That is why the intensities in Fig.~7 can be considered only
as a zero-order approximation, an upper limit.

\begin{figure}
 \centerline{\includegraphics[width=88mm]{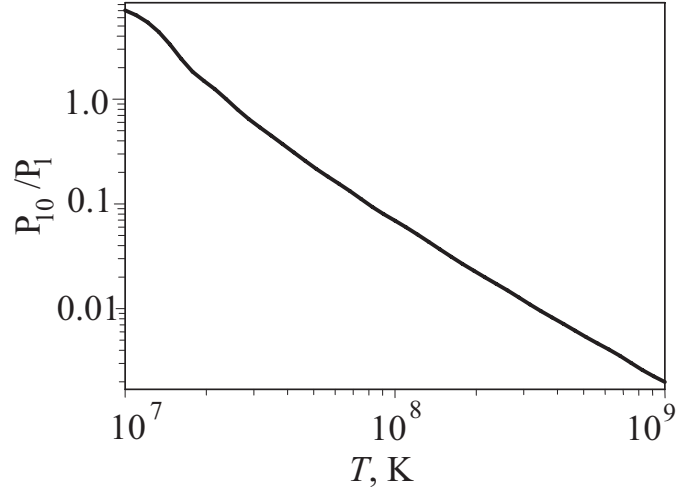}}
 \caption{Ratio of powers of  the satellite line
 \ion{Fe}{xxv}, 1.790 \AA, (No.~10 in Table 2)
 to the resonance line \ion{Fe}{xxvi}, 1.780 \AA,
 (No.~1 in Table 2) as a function of temperature,
 computed using data by Mewe et al. (1982).}
 \label{fig-power-ratio}
\end{figure}

Meanwhile,
Dubau et al. (1981) and
Pike et al. (1996)
show that the \ion{Fe}{xxv} satellites can be very useful
despite they are faint.
They are formed at 1.79 \AA  \,\, due to transitions $1s2l - 2p2l$ ($l=s,p$).
The ratio of \ion{Fe}{xxv}  satellites to  \ion{Fe}{xxvi}  Ly$_\alpha$ lines
(1.780 \AA) is independent of ionization balance but is a function of
temperature
(Fig.~\ref{fig-power-ratio}).
The role of the low-temperature satellite line \ion{Fe}{xxv}
relative to the the super-hot resonance line \ion{Fe}{xxvi}
decreases significantly in increasing $T$.

Observations provide us intensities of spectral lines. Their
interpretation is not easy, because it depends on the $DEM(T)$
(Eq.~\ref{eq-intensity}). Very often the emission source is
assumed to be isothermal. Then the intensity ratio equals
the power ratio:
\[
\frac{I_{10}}{I_1}=\frac{P_{10}}{P_1}=f(T) \, .
\]
Knowing the intensity ratio from observations, we can
determine the plasma temperature using
Fig.~\ref{fig-power-ratio}.

\begin{figure}
 \centerline{\includegraphics[width=88mm]{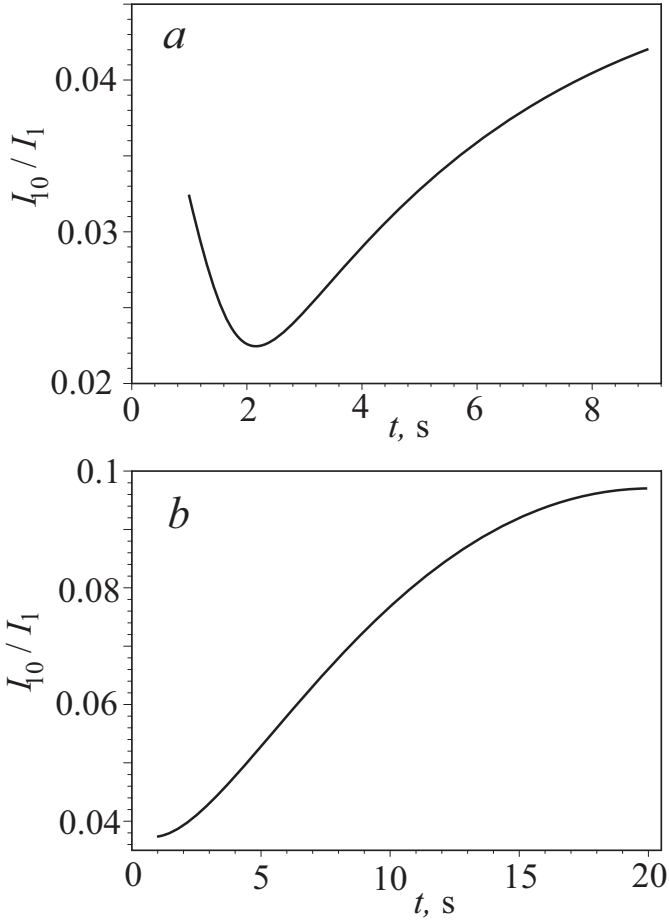}}
 \caption{Ratio of intensities of  the satellite line
 \ion{Fe}{xxv}, 1.790 \AA, (No.~10 in Table 2)
 to the resonance line \ion{Fe}{xxvi}, 1.780 \AA,
 (No.~1 in Table 2) (a) for model 1 and
 (b) for model 2.}
 \label{fig-intensity-ratio}
\end{figure}

Let us consider the results obtained in the frame of this
isothermal approach on the basis of our theoretical intensities.
Figure~\ref{fig-intensity-ratio} demonstrates computed
\ion{Fe}{xxv}/\ion{Fe}{xxvi} intensity ratio for models 1 (a)
and 2 (b). The plasma is very hot, which leads to a low
intensity ratio $\approx 0.022- 0.042$ for the first model. Using
Fig.~\ref{fig-power-ratio}, we obtain the temperature $T\approx
1.5\times 10^8$. It is a good approximation, but the difficulty
consists in the faint intensity $I_{10}$
(fig.~\ref{fig-intensity}, grey curves): $I_{10}\la
0.1~\mathrm{photons}\,\mathrm{cm}^{-2}\,\mathrm{s}^{-1}$.

In model 2, the plasma is relatively cooler and the ratio
$I_{10}/I_{1}$ is higher: $0.04 - 0.1$. It corresponds to
the temperature change from $1.5\times 10^8$~K at the first few seconds
to $5\times 10^7$~K at $t\approx 20$~s. At the very beginning
of a flare, it is very difficult to register $I_{10}$, because its
intensity is less than
$0.01~\mathrm{photons}\,\mathrm{cm}^{-2}\,\mathrm{s}^{-1}$. For
example, Yohkoh observed these lines had integration
time periods that were too long, from 24~seconds to 20~minutes (Pike et al. 1996).
However,
the inferred temperature at 20~s is about $5\times 10^7$~K and
decreases in time. Thus, we do not obtain temperatures that are more than
$10^8$~K, even when they are present in the current layer. Theoretically,
we can observe these lines, but we need a better telescope.
Rough estimations are the following. To detect at least one photon
per second from the satellite line with the intensity of about
$I_{10} \sim
0.01~\mathrm{photons}\,\mathrm{cm}^{-2}\,\mathrm{s}^{-1}$, we need
the effective area of a telescope of about 100 cm$^2$. This is
problematic. We note, however, that even a comparison between data of
Yohkoh and a new instrument having better (not the best)
characteristics (effective area, sensitivity of detectors, etc.)
can give the right trend for detecting the higher temperatures in
flares: the better characteristics, the higher inferred
temperatures in large flares.

Thus, observation of {\it multitemperature and
nonequilibrium effects in flare plasma} are difficult at present, because they
are very small. Sometimes, they are not distinguished from
instrumental effects (e.g. Holman et al. 2011, Kontar et al.
2011). To eliminate this ambiguity or because of
different aims of the work, the majority of observers study
mean plasma characteristics, which are much more reliable. For
example, Phillips et al. (2006)  mostly analyze the flare decay
phase, when the temperature is slowly varying ``to minimize
instrumental problems with high count rates and effects associated
with multitemperature and nonthermal spectral components''. The
authors, however, note that the
isothermal model gives poor spectral fits near the flare maximum.
They write that it can
be due to deviation of the emitting plasma from the isothermal state.
Thus, we can sometimes  see evidence of the
multitemperature nonequilibrium plasma, which should be studied
very carefully.

A more accurate approach than an isothermal one is described
by Doschek \& Feldman (1987). They determine temperature structure
of the radiation source using 16 spectral lines of ions
\ion{S}{xv}, \ion{Ca}{xix}, \ion{Ca}{xix}, \ion{Fe}{xx},
\ion{Fe}{xxv}, \ion{Fe}{xxvi}, and \ion{Ni}{xxvii}. The analysis
demonstrates the possible existence of a super-hot component that
extends up to several hundred million degrees.
The fraction of the emission measure of the super-hot component
to the emission measure of the lower
temperature component is low, $
\la 0.05 - 0.1$. This conclusion may be considered as an indirect
confirmation for a SHTCL.

\section{Conclusions}

We describe an analytical model of high-temperature
turbulent-current layer. It is shown that the thickness of the
layer is very small, of about a few tens of centimeters, whereas the
width and the length are about $10^9$~cm. Its temperature can
be as high as $3\times 10^8$~K. This model allows us to compute
the temperature distribution in the vicinity of the SHTCL at the
beginning of a solar flare. The model considers only
heat conduction, the main effect. Its analysis shows that the
classical collisional approach is not valid under conditions of
solar flare, because we deal with processes that are too fast and
wave fronts that are too steep. For a more accurate description, we
considered the heat-flux relaxation (Oreshina \& Somov
2011). It significantly changes the form of the thermal wave,
slows down the front velocity, and decreases the heat input in a
magnetic tube during its contact with the SHTCL.

Using the obtained temperature distributions in the vicinity of the layer,
we compute
the observable characteristics of the plasma in solar flares:
differential and integral emission measure and intensity of some
spectral lines.
These values are computed for two sets of input parameters for
the SHTCL models.
The difference between the results demonstrates the diagnostic
capabilities of our method, which
creates a theoretical basis for determining characteristic
values of the current layer from future observations.

Particular attention is given to lines of high-charged ions
\ion{Fe}{xxvi} ($\lambda=1.780$~\AA),
\ion{Fe}{xxvi} ($\lambda=1.510$~\AA),  and
\ion{Ni}{xxvii} ($\lambda=1.590$~\AA), which
are formed at $T\gid 10^8$~K.
Their observations can be very useful
for detecting SHTCLs
at the pre-impulsive stage, when the heat conduction is
usually a dominant process.
The later effects, such as plasma heating by accelerated particles,
magnetic traps, chromospheric evaporation, etc., can significantly
complicate the data and their interpretation.
However, the difficulty consists in the weak intensities of spectral lines
at first seconds.

The ratio of the \ion{Fe}{xxv} satellite line ($\lambda=1.790$~\AA) to
the \ion{Fe}{xxvi} resonance line ($\lambda=1.780$~\AA) is of
particular interest, because it does not depend on the ionization
balance. This ratio is often used to determine the plasma
temperature using an isothermal approach. We show that this
technique gives high temperatures ($>10^8$~K) only at
the very beginning of a flare (about seconds) and then the
inferred temperature rapidly decreases. Meanwhile, early
observations are difficult because of a very weak \ion{Fe}{xxv}
satellite line. We expect one photon per cm$^2$ during a few tens
of seconds. The integration time of actual detectors is about some
tens of seconds and more. It can be one of the causes why the
inferred temperatures do not exceed $5\times 10^7$~K, whereas
plasma with $T\ga 10^8$~K is present in a flare.

The other reason why these temperatures are not reported is that
the isothermal approach for interpreting observations (e.g.,
Phillips 2004, Phillips et al. 2006, Holman et al. 2011, Kontar et
al. 2011) gives volume-averaged temperatures. It is
a good method to determine the mean characteristics of flare
plasma. Thus, it gives maximum temperatures of about $30-50$~MK.
However, Phillips (2004), for example, writes at the end of Sect.~4:
``An isothermal plasma is implicit in equations (1), (2),
and (4) and in the discussion so far, but more realistically the
plasma has a nonthermal temperature structure, describable by a
differential emission measure (DEM), $\phi (T_e) = N_e^2\,
dV/dT_e$... .''

We emphasize that the aim of our work is to find the evidence of
the SHTCLs. It is {\it an effect of the second order as compared
to the mean plasma characteristics}, because  a percentage of the
super-hot ($T\ga 10^8$~K) plasma in flares is small. As we can see
from Fig.~4 of our paper, the differential emission measure
has a main maximum at the temperatures of about 10-30 MK. That is, there
is much more plasma in flares at the temperature of 10-30 MK.
These values agree well with the isothermal estimations. Only a
small peak of $DEM(T)$ at $T\ga 10^8$~K is due to the SHTCL.
Finding it from observations is not an easy task. The isothermal
approach for interpreting observations is very fruitful to
determine the general, mean characteristics of the flare plasma,
but it is not appropriate for our purposes. We have to consider
namely {\it multitemperature and nonequilibrium effects.}

Therefore, for the SHTCL diagnostic, the observations in the X-ray
range of 1.5--1.8~\AA ~with high spectral resolution
(better than 0.01~\AA) and high temporal resolution (seconds)
are needed.
It is also very important to interpret the observations using a
multitemperature approach.

\begin{acknowledgements}
The authors thank an anonymous referee for interesting questions
and comments, which have improved the text.
\end{acknowledgements}


\appendix

\section{Length of magnetic-field tubes}

The maximal length of magnetic-field
tubes ($1.9\times 10^{10}$~cm) is estimated by the following manner.
Let us consider a Bastille Day Flare as an example.
A magnetic-charges model of this flare was described by
Oreshina \& Somov  (2009).

\begin{figure}
\centerline{\includegraphics[width=88mm]{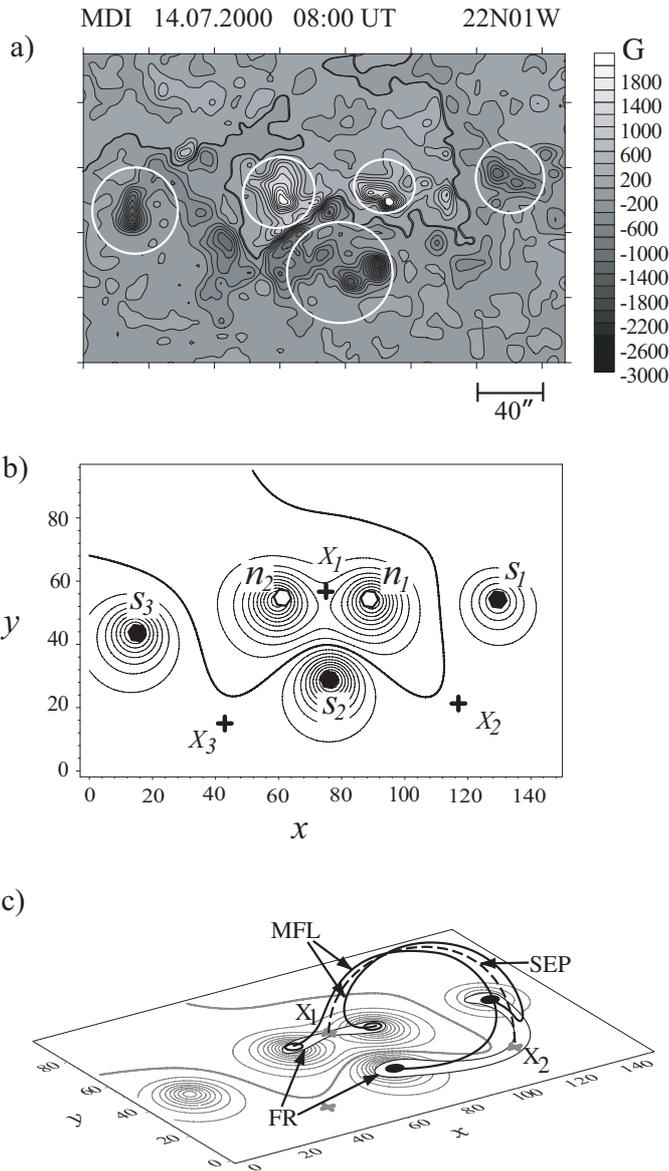}}
\caption{(a) MDI/SOHO magnetograms for AR NOAA 9077;
the magnetic field strength on the isolines is shown in gausses.
(b) Model magnetograms for the same AR with the same isolines:
n1 and n2 are the north-polarity sources;
s1, s2, and s3 are the south-polarity sources,
and X1, X2, and X3 are the null points.
(c) Separator X1X3 (SEP, dashed black line),
magnetic field lines in its vicinity (MFL, black lines),
and flare ribbons (FR). }
\label{fig-model}
\end{figure}

Fig.~\ref{fig-model} presents MDI/SOHO magnetograms for AR NOAA 9077,
model magnetograms for the same AR, and a separator.
 The energy release could be explained by reconnection on two separators:
one connects the null points X1 and X2,
while the other connects the null points X1 and X3 in the plane of the charges.

Both separators have approximately the same length.
As it can be seen from the figure, the distance between the endpoints of
each separator is about 56 pixels of MDI, that corresponds to $8.2\times 10^9$~cm.
In the first approximation, a separator can be considered as a semicircle.
Hence, its top is situated at the height  $R=4.1\times 10^9$~cm.
The distance along the separator from the top to the photosphere is
$L_s=\pi R\, /2=6.4\times 10^9$~cm.

The peculiarity of the separator is that it separates different
magnetic fluxes. Most of the magnetic-field lines are spread from it and
reach the photosphere far from its endpoints.
Thus, a heat wave going from the current layer
to the photosphere along a field line overcomes the distance much greater
than the length of a separator (Fig.~\ref{fig-model}).
It is illustrated also, for example, in the book by
Somov (2006b, fig.~3.7, p.~58).

Thus, the path of heat-wave travelling from the current layer
to the photosphere can be 2-3 times greater than the half-length of the
separator, it can achieve the value $l = L_s\times 3=1.9\times 10^{10}$~cm.

\label{lastpage}

\end{document}